\documentclass[11pt]{article}
\usepackage{amssymb}
\usepackage{graphics}
\usepackage{epsfig}
\usepackage{a4wide}

\begin{document}
\newcommand{\eezzz}{$e^+e^- \to Z^0 Z^0 Z^0$ }
\newcommand{\eezzzg}{$e^+e^- \to Z^0 Z^0 Z^0\gamma$ }

\title{Complete one-loop electroweak corrections to $ZZZ$ production at the ILC}
\author{ Su Ji-Juan, Ma Wen-Gan, Zhang Ren-You, Wang Shao-Ming, and Guo Lei \\
{\small Department of Modern Physics, University of Science and Technology}\\
{\small of China (USTC), Hefei, Anhui 230027, P.R.China}  }

\date{}
\maketitle \vskip 15mm
\begin{abstract}
We study the complete ${\cal O}(\alpha_{ew})$ electroweak (EW)
corrections to the production of three $Z^0$-bosons in the the
framework of the standard model(SM) at the ILC. The leading order
and the EW next-to-leading order corrected cross sections are
presented, and their dependence on the colliding energy $\sqrt{s}$
and Higgs-boson mass $m_{H}$ is analyzed. We investigate also the LO
and one-loop EW corrected distributions of the transverse momentum
of final $Z^0$ boson, and the invariant mass of $Z^0Z^0$-pair. Our
numerical results show that the EW one-loop correction generally
suppresses the tree-level cross section, and the relative correction
with $m_{H}=120~GeV(150~GeV)$ varies between $-15.8\%(-13.9\%)$ and
$-7.5\%(-6.2\%)$ when $\sqrt{s}$ goes up from $350~GeV$ to $1~TeV$.
\end{abstract}


\vfill \eject

\baselineskip=0.32in

\renewcommand{\theequation}{\arabic{section}.\arabic{equation}}
\renewcommand{\thesection}{\Roman{section}.}
\newcommand{\nb}{\nonumber}

\newcommand{\Dir}{\kern -6.4pt\Big{/}}
\newcommand{\Dirin}{\kern -10.4pt\Big{/}\kern 4.4pt}
\newcommand{\DDir}{\kern -7.6pt\Big{/}}
\newcommand{\DGir}{\kern -6.0pt\Big{/}}

\makeatletter      
\@addtoreset{equation}{section}
\makeatother       

\section{Introduction}
\par
To discover the signature of new physics beyond the standard
model(SM)\cite{sm1,sm2} is one of the main goals for the
forthcoming collider experiments. The precision measurements of
the trilinear gauge-boson couplings are helpful for verification
of non-abelian gauge structure, and the investigation of the
quartic gauge-boson couplings can either confirm the symmetry
breaking mechanism or present the direct test on the new physics
beyond the SM\cite{TaoHan}. The direct study of quartic
gauge-boson couplings requires the investigations of the processes
involving at least three external gauge-bosons. In
Refs.\cite{Ohnemus,Campbell,Lazopoulos,Hankele,Binoth}, the
precise predictions for the $VVV$ productions at hadron colliders
were provided. It shows that the QCD corrections increase the
$WWZ$ cross section at the LHC by more than $70\%$, and the QCD
corrections to $ZZZ$ production at the LHC increase the LO cross
section by about $50\%$ \cite{Lazopoulos}. Thus, any quantitative
measurement of the concerned gauge couplings will have to take QCD
corrections into account,

\par
Due to heavy backgrounds, the precise measurement at a hadron
collider is more difficult than at linear collider. The proposed
International Linear Collider (ILC) by the particle physics
community will be built with the entire colliding energy in the
range of $200~GeV <\sqrt{s}<500~GeV$ and an integrated luminosity
of around $500~(fb)^{-1}$ in four years. The machine should be
upgradeable to $\sqrt{s}\sim 1~TeV$ with an integrated luminosity
of $1~(ab)^{-1}$ in three years\cite{ILC}. Among all the ILC
physics goals, the verification of gauge theory in the SM and
finding the evidence of new physics via experimental measurement
of the electroweak gauge boson couplings are crucial tasks too.
The measurement will be able to be improved considerably at ILC
compared with at Fermilab Tevatron and CERN LHC, and therefore a
precise understand of the SM phenomenology at ILC to at least
one-loop order is necessary\cite{Eboli}. Without the accurate
theoretical predictions and reliable error estimates for important
observables at ILC, it is impossible to interpret experimental
data properly.

\par
The process of $ZZZ$ production with the subsequential leptonic
decays of vector bosons at ILC is not only an important process as
a background for various new physics processes, but also possible
to provide further tests for the quadrilinear gauge boson
couplings, including the four gauge boson coupling, such as
between $ZZZZ$, which does not exist at tree-level in the SM,
because this kind of quadrilinear couplings would induce
deviations from the SM predicted observables\cite{eehvv}.

\par
In this paper we present the calculations of the cross sections
for the process \eezzz at the leading order(LO) and involving
complete electroweak (EW) one-loop (${\cal O}(\alpha_{ew})$)
corrections. The paper is organized as follows: In the next
section we present the calculation descriptions for the tree-level
process \eezzz. The calculation of the electroweak corrections at
one-loop level is reported in section III. The numerical results
and discussions are given in section IV. In the last section we
give a short summary.

\vskip 10mm
\section{Leading-order \eezzz process }
\par
\begin{figure*}
\begin{center}
\includegraphics*[125pt,375pt][560pt,550pt]{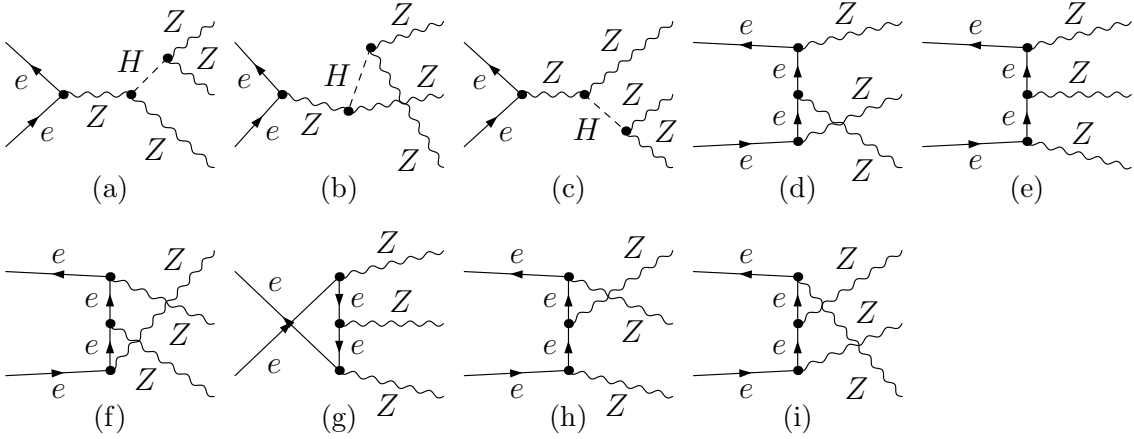}
\caption{\label{fig1} The generic tree-level Feynman diagrams for
\eezzz process. }
\end{center}
\end{figure*}

In the calculations of the tree-level and one-loop level cross
sections for the \eezzz process, we use the 't Hooft-Feynman
gauge. The analytically calculation of the leading order cross
section for \eezzz process is presented in this section. We
describe the lowest order \eezzz process adopted for evaluating
the cross section as
\begin{equation}
\label{process} e^+(p_1)+e^-(p_2) \to Z^0(p_3)+ Z^0(p_4)+Z^0(p_5),
\end{equation}
where $p_i~(i=1-5)$ label the four-momenta of incoming positron,
electron and outgoing $Z^0$-bosons, respectively. Since the mass
of electron/positron is negligible comparing with the colliding
energy and the Yukawa coupling strength between Higgs/Goldstone
and fermions is proportional to the fermion mass, in our work we
ignore the contributions of the Feynman diagrams involving the
couplings of $H^0-e^+-e^-$ and $G^0-e^+-e^-$. We depict the
tree-level Feynman diagrams contributing to the cross section of
the production process of three $Z^0$-bosons at the ILC in
Fig.\ref{fig1}. There we have 9 generic tree-level diagrams for
the process \eezzz in the framework of the SM. All the Born-level
diagrams can be grouped in two different topologies.
Figs.\ref{fig1}(a-c) belong to s-channel, Figs.\ref{fig1}(d-i) are
grouped in t(u)-channel.
The differential cross section for the process \eezzz at the
tree-level is then obtained as
\begin{eqnarray} \label{cross}
d\sigma_{LO} =\frac{1}{3!} \frac{1}{4} \sum_{spin}|{\cal
M}_{LO}|^2 d\Phi_3 ,
\end{eqnarray}
where ${\cal M}_{LO}$ is the amplitude of all the tree-level
diagrams in Fig.\ref{fig1}. The factors $\frac{1}{3!}$ and
$\frac{1}{4}$ are due to three identical final $Z^0$-bosons and
spin-averaging of the initial particles, respectively. The summation
in Eq.(\ref{cross}) is taken over the spins of the initial and final
particles, and $d\Phi_3$ is the three-particle phase space element
defined as
\begin{eqnarray}
d\Phi_3=\delta^{(4)} \left( p_1+p_2-\sum_{i=3}^5 p_i \right)
\prod_{j=3}^5 \frac{d^3 \textbf{\textsl{p}}_j}{(2 \pi)^3 2 E_j}.
\end{eqnarray}

\vskip 10mm
\section{Electroweak (${\cal O}(\alpha_{ew})$) corrections }
\par
The ${\cal O}(\alpha_{ew})$ order electroweak corrections to the
Born-level \eezzz process consist of two parts, i.e.,
\begin{itemize}
\item The virtual contributions to the leading order
process \eezzz from the electroweak one-loop and their corresponding
counterterm diagrams;

\item The contribution from the real photon emission process
\eezzz$\gamma$. The soft photon emission process \eezzz($\gamma$)
consists IR singularities, which will be cancelled by the IR
singularities in the contributions of the one-loop diagrams. There
is no collinear IR singularity since we keep the nonzero mass of
electron(positron);
\end{itemize}
In the following subsections, we describe in detail the
calculation procedure and discuss the calculation of each
contribution part.

\vskip 10mm
\subsection{ Virtual corrections }
\par
\begin{figure}[htbp]
\vspace*{-0.3cm} \centering
\includegraphics*[125pt,295pt][570pt,635pt]{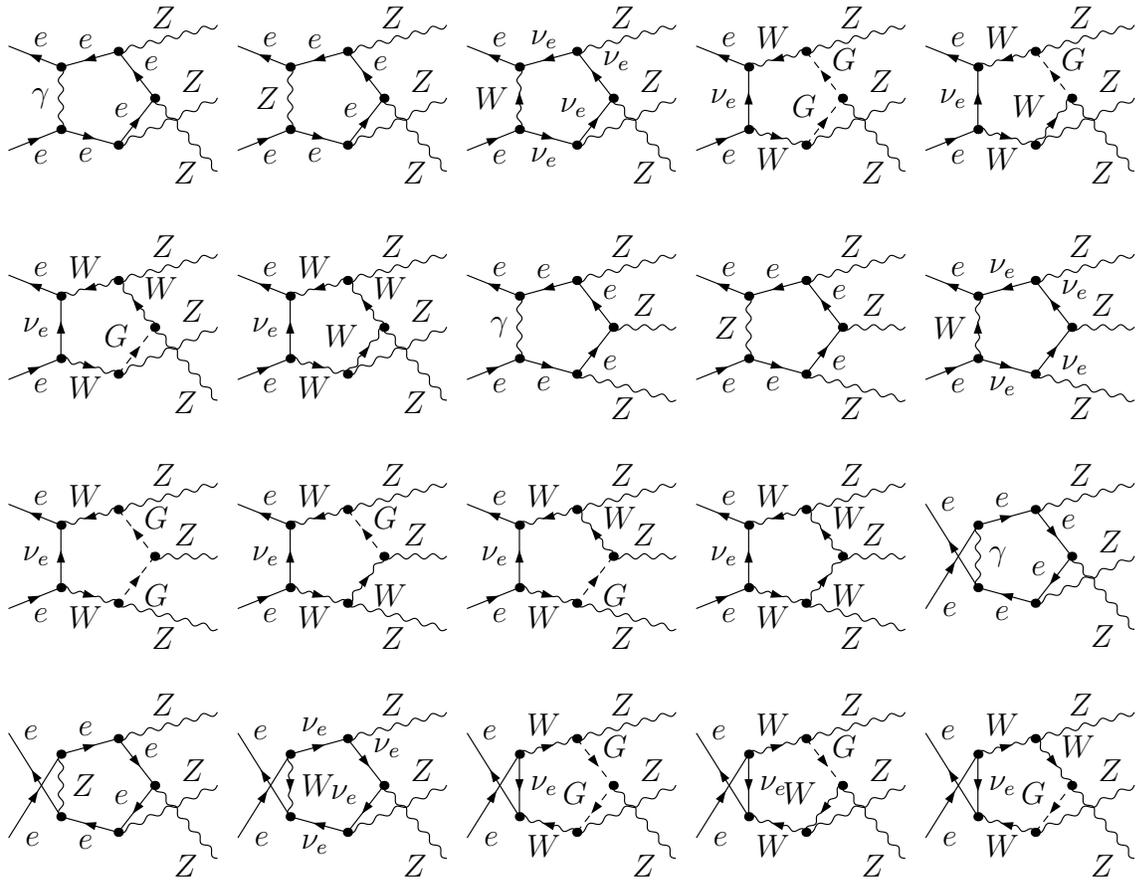}
\vspace*{-0.3cm} \centering \caption{\label{fig2} Some
representative pentagon Feynman diagrams for the process \eezzz.}
\end{figure}

There are totally 2313 electroweak one-loop and corresponding
counterterm Feynman diagrams being taken into account in our
calculation, and they can be classified into self-energy,
triangle, box, pentagon and counterterm diagrams. We depict some
representative samples among 66 pentagon diagrams in
Fig.\ref{fig2}. In our calculation of the electroweak (${\cal
O}(\alpha_{ew})$) corrections to \eezzz process, all the one-loop
Feynman diagrams and their relevant amplitudes are created by
using $FeynArts~3.3$\cite{fey}, and the Feynman amplitudes are
subsequently implemented by applying FormCalc5.3
programs\cite{formloop} and our in-house routines. The electroweak
one-loop amplitude involves five point tensor integrals up to rank
4. The numerical calculation of the integral functions($n \leq 4$)
are implemented by using the expressions presented in
Refs.\cite{OneTwoThree,Four}. We use our independent Fortran
subroutines following the expressions for the scalar and tensor
five-point integrals in Ref.\cite{Five}, and find agreement with
LoopTools2.2\cite{formloop}. The Grace2.2.1 program\cite{Grace} is
used to accomplish five-body phase-space integration for hard
photon radiation process \eezzzg.

\par
The total unrenormalized amplitude corresponding to all the
one-loop Feynman diagrams contains both ultraviolet (UV) and
infrared (IR) singularities. We regulate all singularities
adopting dimensional regularization(DR) scheme \cite{DR} where the
dimensions of spinor and space-time manifolds are extended to $D =
4 - 2 \epsilon$. The relevant fields are renormalized by taking
the on-mass-shell (OMS) scheme \cite{COMS scheme, Denner}. The IR
singularity is regularized by introducing a infinitesimal
fictitious mass $m_{\gamma}$. All the tensor coefficients of the
one-loop integrals can be calculated by using the reduction
formulae presented in Refs.\cite{Five,Passarino}. As we expect,
the UV divergence contributed by virtual one-loop diagrams can be
cancelled by that contributed from the counterterms exactly both
analytically and numerically.

\vskip 10mm
\subsection{ Real photon emission process \eezzzg }
\par
In our calculation for one-loop diagrams, there exists soft IR
divergence. In order to get an IR finite cross section for \eezzz
up to the order of ${\cal O}(\alpha_{ew}^4)$, we should consider
the ${\cal O}(\alpha_{ew})$ corrections to \eezzz process due to
real photon emission. The soft IR divergence in virtual photonic
corrections for the process \eezzz can be exactly cancelled by
adding the real photonic bremsstrahlung corrections to this
process in the soft photon limit. In the real photon emission
process
\begin{eqnarray}
\label{realphoton}
 e^+(p_1)+e^-(p_2) \to Z^0(p_3)+Z^0(p_4)+Z^0(p_5)+\gamma(p_6),
\end{eqnarray}
a real photon radiates from the electron/positron, and can be soft
or hard. The general phase-space-slicing (PSS) method \cite{PSS}
is adopted to isolate the soft photon emission singularity part in
the real photon emission process \eezzzg, and the cross section of
the real photon emission process ($\ref{realphoton}$) is
decomposed into soft and hard terms
\begin{equation}
\Delta\sigma_{{real}}=\Delta\sigma_{{S}}+\Delta\sigma_{{H}}=
\sigma_{LO}(\delta_{{S}}+\delta_{{H}}).
\end{equation}
where the 'soft' and 'hard' describe the energy nature of the
radiated photon. The energy $E_{6}$ of the radiated photon in the
center of mass system(c.m.s.) frame is considered soft if $E_{6}
\leq \Delta E$, and hard if $E_{6} > \Delta E$, respectively. Then
both $\sigma_{{S}}$ and $\sigma_{{H}}$ should depend on the
arbitrary soft cutoff $\delta_s\equiv\Delta E/E_{b}$, where $E_b$ is
the electron beam energy in the c.m.s. frame and equals to
$\sqrt{s}/2$, but the total cross section of the real photon
emission process $\sigma_{{real}}$ is cutoff $\Delta E/E_{b}$
independent. Since the soft cutoff $\Delta E/E_{b}$ is taken to be a
small value in our calculations, the terms of order $\Delta E/E_{b}$
can be neglected and the soft contribution can be evaluated by using
the soft photon approximation analytically \cite{COMS scheme,Denner,
soft r approximation}
\begin{eqnarray}
\label{s} {d} \Delta\sigma_{S} = - d \sigma_{{LO}}
\frac{\alpha_{ew}}{2 \pi^2} \int_{|\vec{p}_6| \leq \Delta E}
\frac{d^3 p_6}{2 E_6} \left( \frac{p_1}{p_1 \cdot p_6} -
\frac{p_2}{p_2 \cdot p_6} \right)^2.
\end{eqnarray}

Our calculation demonstrates that the IR singularity in the soft
contribution from Eq.(\ref{s}) is cancelled exactly with that from
the virtual photonic corrections. Therefore, $\Delta\sigma_{v}
+\Delta\sigma_{real}$, the sum of the ${\cal O}(\alpha_{ew}^4)$
cross section corrections from virtual, soft and hard photon
emission contribution parts, is independent of the cutoff value
$\delta_s$. The hard contribution, which is UV and IR finite, is
computed by using the Monte Carlo technique. Finally, the
electroweak corrected cross section for the \eezzz process up to the
order of ${\cal O}(\alpha^4_{ew})$ can be obtained by
\begin{eqnarray}
\sigma_{{tot}}= \sigma_{LO} + \Delta\sigma_{v} + \Delta\sigma_{real}
= \sigma_{LO} \left( 1 + \delta_{tot} \right).
\end{eqnarray}

\vskip 10mm
\subsection{ QED and total ${\cal O}(\alpha_{ew})$ order corrections }
\par
In analyzing the originations of the electroweak corrections, we
split the full ${\cal O}(\alpha_{ew})$ corrections to the process
\eezzz into two parts, the QED correction part and the weak
correction part. Correspondingly we define the total relative
correction as $\delta_{tot}=\delta_{QED}+\delta_{weak}$. The QED
correction part is contributed by the diagrams with virtual photon
in loop, and real photon emission process \eezzzg. For the
counterterms involved in the QED contribution, the
electron/positron wave function renormalization constants include
only photonic contribution. The remainders of the total virtual
electroweak corrections belong to weak correction part. With above
definitions we can express the full one-loop electroweak corrected
total cross section as a summation of several parts.
\begin{eqnarray}
\sigma_{tot}&=& \sigma_{LO} + \Delta\sigma_{v,QED} +
\Delta\sigma_{s} + \Delta\sigma_{h} +
\Delta\sigma_{v,weak} \nb\\
&=& \sigma_{LO} \left( 1 + \delta_{QED}+\delta_{weak} \right)=
\sigma_{LO} \left( 1 + \delta_{tot} \right),
\end{eqnarray}
where $\Delta\sigma_{v}$ and $\Delta\sigma_{s}$ are the cross
section corrections contributed by the virtual electroweak one-loop
diagrams and the soft photon emission process respectively,
$\Delta\sigma_{v,QED}$, $\Delta\sigma_{s}$, $\Delta\sigma_{h}$ and
$\Delta\sigma_{v,weak}$ are the corrections from the virtual QED
contribution, the soft photon emission process, the hard photon
emission process and the virtual weak contribution, separately.
$\delta_{QED}$, $\delta_{weak}$ and $\delta_{tot}$ are the relative
corrections contributed by the QED correction part, the weak
correction part and the total electroweak correction, respectively.

\par
As we mentioned above, there exist both ultraviolet(UV) divergency
and infrared(IR) soft singularity in the contributions of the
electroweak one-loop diagrams for \eezzz process, but no collinear
IR singularity since we keep the nonzero mass of electron/positron
in our calculation of one-loop order corrections. After doing the
renormalization procedure, we verified that the UV singularity is
vanished, and the IR soft divergency appeared in the virtual
correction is cancelled by the the soft photon emission process
\eezzzg.

\par
In order to discuss the origin of the large correction when the
colliding energy is close to the threshold of the production of
three $Z^0$-bosons, we discuss the photonic (QED) corrections and
the genuine total electroweak corrections separately. The QED
corrections comprise two parts: the QED virtual corrections
$\Delta \sigma_{v,QED}$ which contributed by the loop diagrams
with virtual photon exchange in loop and the corresponding QED
parts of the counterterms, and the real photon emission
corrections $\Delta \sigma_{real}$. Therefore, the QED relative
correction $\delta_{QED}$ can be expressed as
\begin{eqnarray}
\delta_{QED} = \delta_{v,QED} + \delta_{real},
\end{eqnarray}
where $\delta_{v,QED} = \Delta \sigma_{v,QED}/\sigma_{LO}$, and
the genuine weak relative correction $\delta_{w}$ can be got from
\begin{eqnarray}
\delta_{w} = \delta_{tot} - \delta_{QED}.
\end{eqnarray}

\vskip 10mm
\section{Numerical results and discussion}
\par
\par
For the numerical calculation, we take the input parameters as
follows\cite{JofP}:
\begin{eqnarray}\label{input}
m_e&=&0.51099892~{\rm MeV},~m_\mu~=~105.658369~{\rm MeV},~m_\tau~=~1776.99~{\rm MeV},\nb\\
m_u&=&66~{\rm MeV},~~~~~~~~~~~~~~m_c~=~1.25~{\rm GeV},~~~~~~~~~~~~~m_t~=~174.2~{\rm GeV},\nb\\
m_d&=&66~{\rm MeV},~~~~~~~~~~~~~~m_s~=~95~{\rm MeV},~~~~~~~~~~~~m_b~=~4.7~{\rm GeV} ,\nb\\
m_W&=&80.403~{\rm GeV},~~~~~~~~~m_Z~=~91.1876~{\rm GeV}.
\end{eqnarray}
There we use the experimental value of W-boson mass as input
parameter, but not the $m_W$ evaluated from $G_{\mu}$ as in
$\alpha_{ew}$ scheme\cite{alpha-scheme}. We take the electric
charge defined in the Thomson limit $\alpha_{ew}(0) = 1/137.036$
and the effective values of the light quark masses ($m_u$ and
$m_d$) which can reproduce the hadronic contribution to the shift
in the fine structure constant $\alpha_{ew}(m_Z^2)$ \cite{DESY}.
As we know that the LEP II experiments provide the lower limit on
the SM Higgs mass as $114.4~ GeV$ at the $95\%$ confidence level
from the results of direct searches for $e^+e^- \to Z^0H^0$
production\cite{lower mH,upper mH}, and the electroweak precision
measurements indicate indirectly the upper bound as $m_H \lesssim
182~GeV$ at the $95\%$ C.L., when the lower limit on $m_H$ is used
in determination of this upper limit\cite{upper mH}. Therefore, in
our numerical evaluation it is reasonable to take the mass of
Higgs-boson being in the range of $115~GeV < m_H < 170~GeV$. Then
we shall not encounter the resonance problem of Higgs-boson during
our calculation.

\par
We checked the correctness of the numerical results of the LO
cross section for process \eezzz, by using Grace2.2.1\cite{Grace}
and FeynArts3.3/FormCalc5.3 \cite{fey,formloop} packages
separately. In adopting Grace2.2.1 and FeynArts3.3/FormCalc5.3
programs, we used both 't Hooft-Feynman and unitary gauges
separately in the calculations of the LO cross section to check
the gauge invariance, and we got coincident numerical results. The
numerical results of the LO cross section for the process \eezzz,
by using 't Hooft-Feynman gauge and taking $\sqrt{s}=500~GeV$ and
the other input parameters shown in Eqs.(\ref{input}), are listed
in Table \ref{tab1}. There it is shown that there is a good
agreement between the numerical results by adopting different
packages.
\begin{table}
\begin{center}
\begin{tabular}{|c|c|c|}
\hline $m_H (GeV)$
& $\sigma_{LO}(fb)$(Grace) & $\sigma_{LO}(fb)$ (FeynArts) \\
\hline 115  & 1.0056(4)  & 1.0055(2)  \\ \hline 120  & 1.0139(4) &
1.0138(2) \\ \hline 150  & 1.0975(4) & 1.0975(2)  \\
\hline 170  & 1.2565(4) & 1.2564(2)  \\\hline
\end{tabular}
\end{center}
\begin{center}
\begin{minipage}{15cm}
\caption{\label{tab1} The numerical results of the LO cross
sections for the process \eezzz by using Grace2.2.1 and
FeynArts3.3/FormCalc5.3 packages separately, and taking the input
parameters as shown in Eqs.(\ref{input}) and $\sqrt{s}=500~GeV$. }
\end{minipage}
\end{center}
\end{table}

\par
In the one-loop calculation, we must set the values of the IR
regulator $m_{\gamma}$, the fictitious photon mass, and soft
cutoff $\delta_s=\Delta E/E_b$ besides the parameters mentioned in
Eqs.(\ref{input}). As we know, the total cross section should have
no relation with these two parameters if the IR divergency does
really vanish. Our numerical results show that the cross section
correction at ${\cal O} (\alpha^4_{ew})$ order
$\Delta\sigma_{tot}=\Delta\sigma_{real}+\Delta\sigma_{v}$ is
invariable within the calculation errors when $m_H=120~GeV$,
$\sqrt{s}=500~GeV$, $\delta_s=10^{-3}$ and the fictitious photon
mass $m_{\gamma}$ varies from $10^{-15}~ GeV$ to $10^{-1}~GeV$.

\par
\begin{figure}
\includegraphics[scale=0.78]{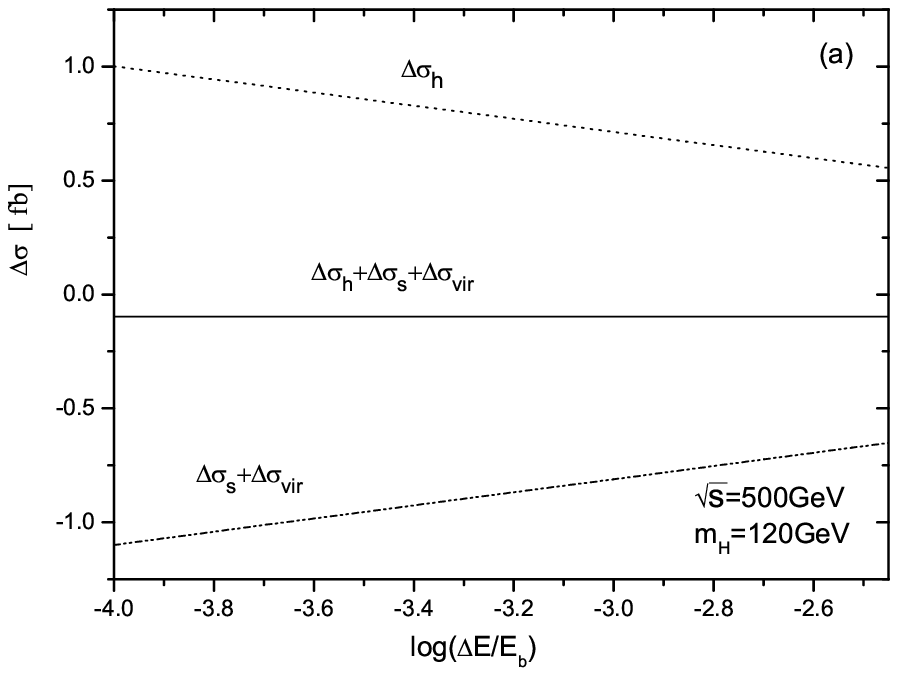}
\includegraphics[scale=0.4]{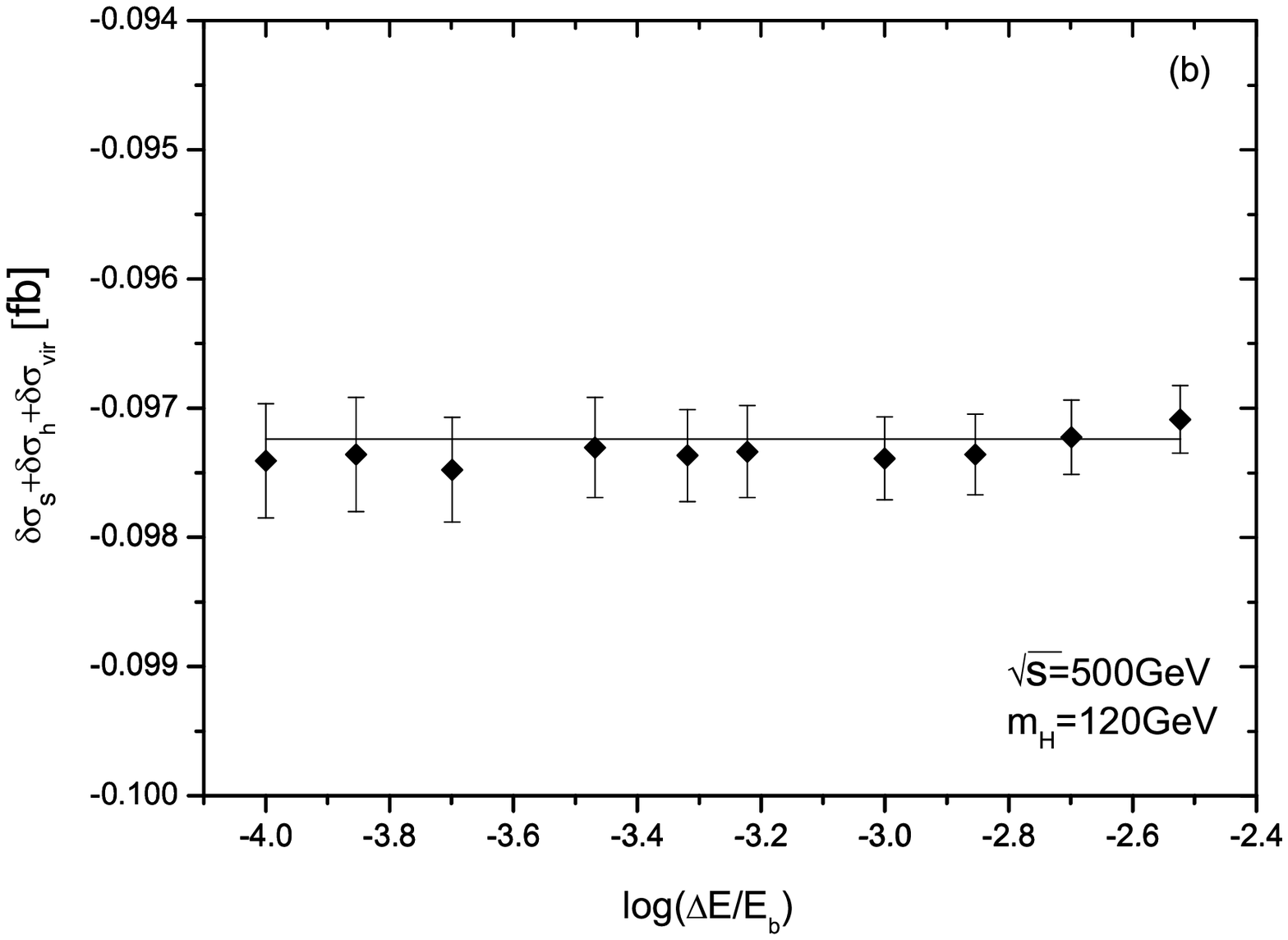}
\caption{\label{fig3} (a) The dependence of the ${\cal
O}(\alpha_{{\rm ew}})$ correction to cross section of \eezzz on
the soft cutoff $\delta_s$ with $m_H=120~GeV$ and
$\sqrt{s}=500~GeV$. (b) the amplified plot of the curve for the
total correction $\Delta\sigma_{tot}$ in Fig.3(a), where it
includes calculation errors.}
\end{figure}

Fig.\ref{fig3}(a) presents a verification of the correctness of our
calculation for process \eezzz including ${\cal O}(\alpha_{ew})$
order corrections. The amplified curve for $\Delta\sigma_{tot}$
including calculation errors is depicted in Fig.\ref{fig3}(b). Both
figures are to show the independence of the total ${\cal
O}(\alpha_{{\rm ew}})$ electroweak correction on the soft cutoff
$\delta_s$, when we take $m_{H} = 120~ {{\rm GeV}}$ and $\sqrt{s} =
500~ {\rm GeV}$. From Fig.\ref{fig3}(b) we can say that the total EW
relative correction $\Delta\sigma_{tot}$ has no relation to the
value of $\delta_s$ within the calculation error range. In the
further calculations, we set $m_{\gamma}=10^{-2}~GeV$ and
$\delta_s=10^{-3}$.

\par
In Table \ref{tab2}, we list some representative numerical results
of the LO and one-loop EW corrected cross sections($\sigma_{LO}$,
$\sigma_{tot}$), the QED and total EW corrections to the cross
sections for \eezzz process($\Delta\sigma_{QED}$,
$\Delta\sigma_{tot}$), and their corresponding relative
corrections($\delta_{QED}$, $\delta_{tot}$) when
$\sqrt{s}=500~GeV$ and the values of Higgs-boson mass are taken to
be $115~GeV$, $150~GeV$ and $170~GeV$ separately. From these data
we can see that the one-loop EW corrections suppress the LO cross
section of the process \eezzz, and the relative corrections are
about minus few percent. We can conclude also the LO and EW
corrected cross sections increase with the increment of $m_H$,
while the absolute total EW correction decreases when the value of
Higgs-boson mass goes up.
\begin{table}
\begin{center}
\begin{tabular}{|c|c|c|c|c|c|c|c|}
\hline $m_H (GeV)$ & $\sigma_{LO}(fb)$ & $\sigma_{tot}(fb)$ &
$\Delta\sigma_{QED}(fb)$ & $\Delta\sigma_{tot}(fb)$ &
$\delta_{QED}(\%)$ & $\delta_{tot}(\%)$  \\
\hline 115  & 1.0055(2) & 0.9159(7) & -0.0451(7) & -0.0896(7) & -4.49(7) & -8.91(7)   \\
\hline 150  & 1.0975(2) & 1.0194(8) & -0.0444(8) & -0.0780(8) & -4.04(7) & -7.11(7)  \\
\hline 170  & 1.2564(2) & 1.1989(9) & -0.0393(8) & -0.0575(9) &
-3.12(7) & -4.58(7)
\\\hline
\end{tabular}
\end{center}
\begin{center}
\begin{minipage}{15cm}
\caption{\label{tab2} The numerical results of $\sigma_{LO}$,
$\sigma_{tot}$, $\Delta\sigma_{QED}$, $\Delta\sigma_{tot}$(in
femto bar), and their corresponding relative EW and QED
corrections($\delta_{tot}$, $\delta_{QED}$) for the process
\eezzz, when $\sqrt{s}=500~GeV$ and
$m_H=115~GeV,~150~GeV,~170~GeV$ respectively. }
\end{minipage}
\end{center}
\end{table}

\par
\begin{figure}
\centering
\includegraphics[scale=0.65]{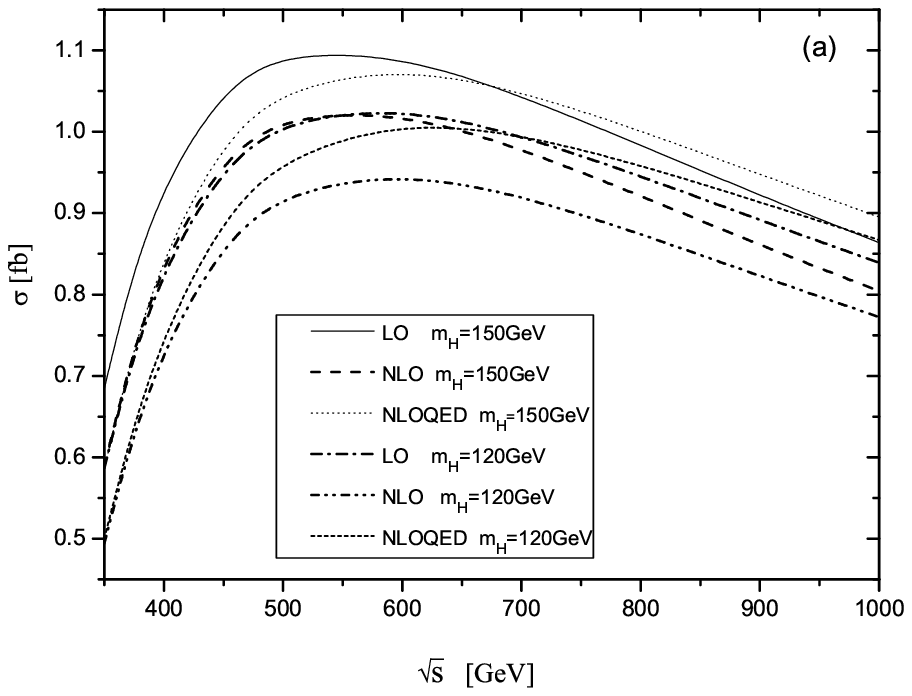}
\includegraphics[scale=0.65]{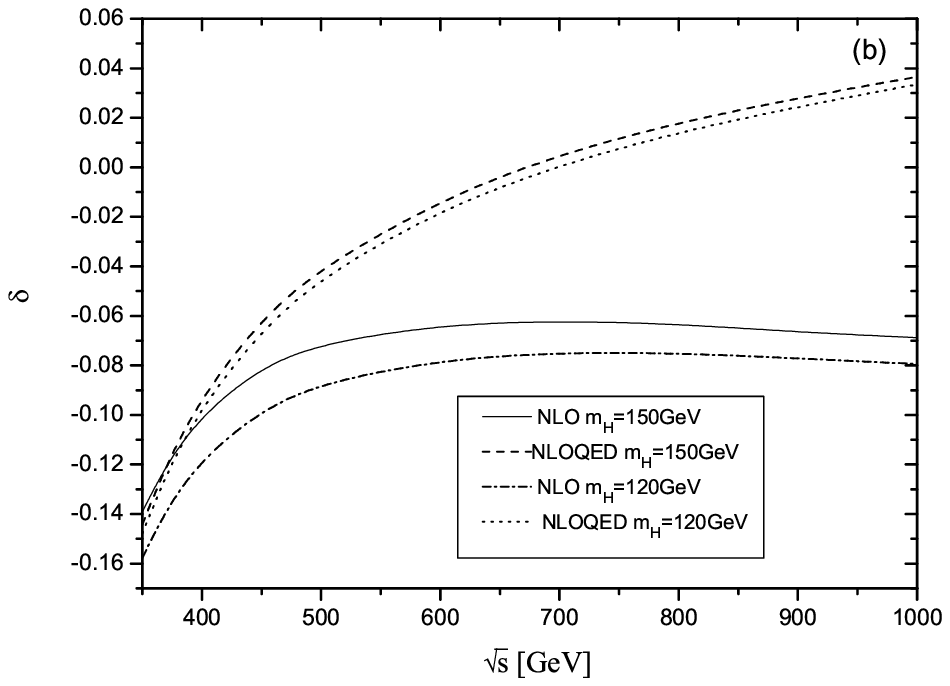}
\caption{\label{fig4} (a) The LO($\sigma_{LO}$), ${\cal
O}(\alpha_{ew})$ EW, QED corrected cross sections($\sigma_{tot}$,
$\sigma_{QED}$) for the process \eezzz as the functions of colliding
energy $\sqrt{s}$ with $m_H=120~GeV$, $150~GeV$ separately. (b) The
corresponding relative EW, QED relative corrections($\delta_{tot}$,
$\delta_{QED}$) versus $\sqrt{s}$.}
\end{figure}

The numerical results of the LO, ${\cal O}(\alpha_{ew})$ EW, QED
corrected cross sections($\sigma_{LO}$, $\sigma_{tot}$,
$\sigma_{QED}$) and the total relative EW, QED
corrections($\delta_{tot}$, $\delta_{QED}$) for the process \eezzz
with $m_H=120~GeV,~150~GeV$ as the functions of colliding energy
$\sqrt{s}$ are plotted in Figs.\ref{fig4}(a) and (b) respectively.
As indicated in Fig.\ref{fig4}(a), The curves for the cross sections
of $\sigma_{LO}$, $\sigma_{tot}$  and $\sigma_{QED}$ increase
quickly in the $\sqrt{s}$ region of $[350~GeV,~550~GeV]$ and
decrease when $\sqrt{s}>600~GeV$. The two figures show the ${\cal
O}(\alpha_{ew})$ corrections always suppress the corresponding LO
cross sections of process \eezzz in both cases of $m_H=120~GeV$ and
$m_H=150~GeV$ separately, but the QED correction parts can enhance
the LO cross sections when $\sqrt{s}>700~GeV$. We can read out from
Fig.\ref{fig4}(b) that the corresponding total EW relative
corrections for $m_H=120~GeV$ and $150~GeV$ vary in the ranges of
$[-15.8\%,~-7.5\%]$ and $[-13.9\%,~-6.2\%]$ respectively, when
$\sqrt{s}$ runs from $350~GeV$ to $1~TeV$. We can see also from
these two plots that in the colliding energy $\sqrt{s}$ region near
the threshold of the production of three $Z^0$-bosons, the main
contribution to the total EW relative correction($\delta_{tot}$)
comes from the QED correction part. That is due to the Coulomb
singularity effect coming from the instantaneous photon exchange in
loops which has a small spatial momentum.

\par
\begin{figure}
\centering
\includegraphics[scale=0.5]{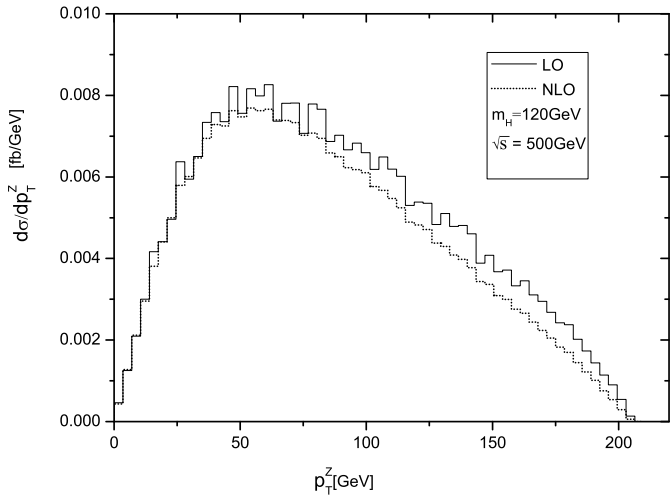}
\caption{\label{fig5} The distributions of the transverse momenta of
$Z^0$-bosons ($P_T^Z$) at the LO and up to EW one-loop order with
$m_H=120~GeV$ and $\sqrt{s}=500~GeV$. }
\end{figure}

We present the distributions of the transverse momenta of final
$Z^0$-bosons at leading order and up to one-loop order,
$d\sigma_{LO}/dp_T^{Z}$, $d\sigma_{NLO}/dp_T^{Z}$, when
$m_H=120~GeV$ and $\sqrt{s}=500~GeV$ in Fig.\ref{fig5}. There we
pick the $p_T^Z$ of each of the three $Z^0$-bosons as an entry in
this histograms, then the final result of  the differential cross
section is obtained by multiplying factor $1/3$. In this figure we
can see that the EW one-loop correction suppresses obviously the
LO differential cross section $d\sigma_{LO}/dp_T^{Z}$ when $p_T^Z
> 50~GeV$, but the EW correction is small when $p_T^Z < 25~GeV$.
It also shows that the EW corrections do not observably change the
LO distribution line-shape of $p_T^Z$, and both the differential
cross sections of $d\sigma_{LO}/dp_T^{Z}$ and
$d\sigma_{NLO}/dp_T^{Z}$ have their maximal values at about
$p_T^{Z}\sim 50~GeV$ respectively.

\par
\begin{figure}
\centering
\includegraphics[scale=0.5]{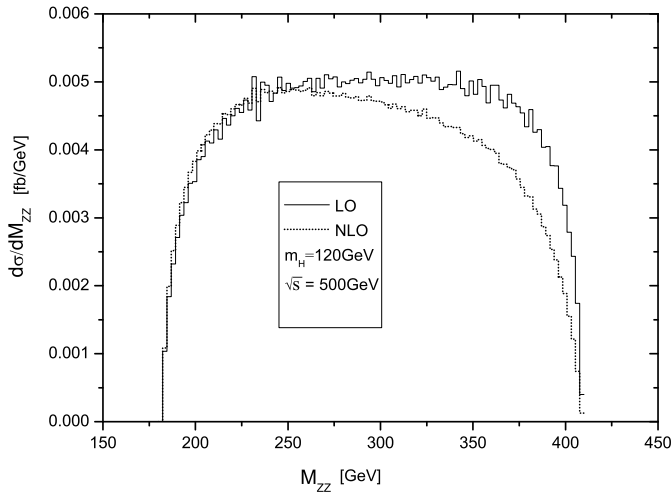}
\caption{\label{fig6} The distributions of the invariant mass of
$Z^0Z^0$-pair($M_{ZZ}$) at the LO and up to EW one-loop order when
$m_H=120~GeV$ and $\sqrt{s}=500~GeV$. }
\end{figure}

We plot the distributions of the invariant mass of $Z^0Z^0$-pair,
denoted as $M_{ZZ}$, at the LO and up to EW one-loop order with
$m_H=120~GeV$ and $\sqrt{s}=500~GeV$ in Fig.\ref{fig6}. Here we
can see that the EW correction slightly enhances the LO
differential cross section when $M_{ZZ} < 250~GeV$, but suppresses
$d\sigma_{LO}/dM_{ZZ}$ obviously when $M_{ZZ}
> 250~GeV$. The suppression of $d\sigma_{LO}/dM_{ZZ}$ is due to
the contribution from the hard photon emission process \eezzzg at
the ${\cal O}(\alpha_{ew}^4)$ order, in which the momentum balance
between the sum of the momenta of three $Z^0$-bosons and that of the
radiated hard photon will reduce the value of invariant mass
$M_{ZZ}$ and change the line-shape in the range with large $M_{ZZ}$.

\vskip 10mm
\section{Summary}
In this paper we describe the impact of the complete one-loop EW
corrections to the scattering process \eezzz in the SM. This channel
can be used to measure the quartic vector boson coupling at ILC. We
investigate the dependence of the LO, ${\cal O}(\alpha_{ew})$ EW and
QED corrected cross sections on colliding energy $\sqrt{s}$ and
Higgs-boson mass, and present the LO and EW one-loop corrected
distributions of the transverse momenta of final $Z^0$-bosons and
the LO and EW corrected differential cross sections of invariant
mass of $Z^0Z^0$-pair. To see the origin of some of the large
corrections clearly, we calculate the QED and genuine weak
corrections separately. We conclude that both the Born cross section
and the EW corrected cross section for \eezzz process are sensitive
to the Higgs boson mass in the range of $115~ GeV < m_H < 170~ GeV$.
We find the ${\cal O}(\alpha_{ew})$ corrections generally suppress
the LO cross section, the LO distribution of the momenta of
$Z^0$-bosons and the LO differential cross sections of invariant
mass of $Z^0Z^0$-pair for process \eezzz. Our numerical results show
that when $m_{H}=120~GeV(150~GeV)$ and the colliding energy goes up
from $350~GeV$ to $1~TeV$, the relative EW correction varies from
$-15.8\%(-13.9\%)$ to $-7.5\%(-6.2\%)$.

\vskip 6mm
\par
\noindent{\large\bf Acknowledgments:} This work was supported in
part by the National Natural Science Foundation of China,
Specialized Research Fund for the Doctoral Program of Higher
Education(SRFDP) and a special fund sponsored by Chinese Academy
of Sciences.

\end{document}